\documentclass[12pt,preprint]{aastex}

\shorttitle{H~{\footnotesize{I}} Continuum Absorption in the Galactic Plane}
\shortauthors{S.T. Strasser, A.R. Taylor}

\newcommand{\HI}{H~{\footnotesize{I}}~}

\begin{document}
\title{1420 MHz Continuum Absorption toward Extragalactic Sources
       in the Galactic Plane}
\author{S. Strasser} \affil{Department of Astronomy, University of Minnesota, Minneapolis, MN 55455}
\author{A.R. Taylor} \affil{Department of Physics and Astronomy, University of Calgary, Calgary, AB T2N 1N4} 
\email{strasser@astro.umn.edu, russ@ras.ucalgary.ca}

\begin{abstract}
We present a 21-cm emission-absorption study towards extragalactic sources in the Canadian
Galactic Plane Survey (CGPS). We have analyzed \HI spectra towards 437 sources with $S_{\nu} \ge$ 150 mJy, 
giving us a source density of 0.6 sources per square degree at arcminute resolution. We present the 
results of a first analysis of the \HI temperatures, densities, and feature statistics. Particular 
emphasis is placed on 5 features with observed spin temperatures below 40 K.

We find most spin temperatures in the range from 40 K to 300 K. A simple \HI two-component model constrains the 
bulk of the cold component to temperatures ($T_c$) between 40 K and 100 K. $T_c$ peaks in the Perseus arm 
region and clearly drops off with Galactocentric radius, R, beyond that. The \HI density follows this trend, 
ranging from a local value of 0.4 cm$^{-3}$ to less than 0.1 cm$^{-3}$ at R = 20 kpc. We find that 
\HI emission alone on average traces about 75\% of the total \HI column density, as compared to the total
inferred by the emission and absorption. Comparing the neutral hydrogen absorption to CO emission no 
correlation is found in general, but all strong CO emission is accompanied by a visible \HI spectral feature. 
Finally, the number of spectral \HI absorption features per kpc drop off exponentially with increasing R. 
\end{abstract}


\section{INTRODUCTION}

In the last 30 years our knowledge of the interstellar neutral hydrogen (H~{\footnotesize{I}}) in the Galaxy 
has expanded dramatically. Many of the discoveries came from emission-absorption studies towards 
continuum background sources at 1420 MHz. Perhaps the most important realization of the early surveys 
\citep[e.g.][]{c65, htc71, rmlw72} was that the emission line-widths are clearly broader than the 
corresponding absorption, indicating that the emitting gas must be dominated by a much warmer phase than the absorbing
gas. \cite{l75}, \cite{dst78}, \cite{mwkg82} 
analyzed the spin temperature 
($T_s$), and found values from approximately 40 K up to several thousand Kelvin, peaking 
between 80 K and 150 K. Due to the probable mix of cold and warm gas, the physical 
interpretation of $T_s$ has remained difficult. \cite{bw92} applied a two-component model to absorption
and emission data and found cold phase temperatures in agreement with the above spin temperatures. 
However two recent studies \citep{ht02,ht02b,d02}, fitting individual features, find the cold component temperature 
peaking between 40 K and 60 K. \cite{ht02b} also
finds the warm component ranging from roughly 1000 K to 9000 K.
This puts ideas about the ISM existing in distinct stable temperature and density regimes 
\citep[e.g.][]{mo77} into question. We also do not have a firm theoretical or observational basis 
on the lower limit (if such exists) of the cold component temperature, before most of the gas 
becomes molecular. 

Addressing the variation with Galactocentric radius, \cite{kh88} state 
that the supernova-dominated McKee-Ostriker (MO) model \citep{mo77} might describe the inner Galaxy better, 
while the outer, more quiescent, Galaxy is better described by the FGH model \citep{fgh69}.
\cite{dmgg02} and \cite{kjbd02} define the mean opacity, $<\kappa>$, as the optical depth line of the 
sight integral divided by the radial velocity change due to the rotation curve, for radial increments in 
the Galaxy. They found an decrease in $<\kappa>$ with increasing radius in the first and fourth 
quadrant respectively. Overall, however, our knowledge of the radial variation of the ISM phases is still 
quite limited. This is mainly because of the time requirements to observe a large number of sources over a 
wide area of the Galactic plane at the sensitivity needed to get a good source density. New surveys, as those
used by \cite{dmgg02} and \cite{kjbd02}, are starting to give us a much better picture of the \HI states
and the variations with Galactocentric radius.

We present a new emission-absorption study toward compact continuum background sources in 
the Canadian Galactic Plane Survey \citep[CGPS,][]{t03}. We have applied an ``on-off method'' to extract the 
emission brightness temperature and optical depth toward the sources. Due to the high resolution
and sensitivity of the CGPS we are able to study the radial variation of the \HI temperature and density
for a much higher source density in the Galactic Plane than ever before. Section 2 gives an overview
of the CGPS and the on-off method used. In section 3 we present the spin temperatures and the results 
of a simple two-temperature model. Further we discuss the spatial variation of the \HI column density, 
and the correlation with CO. We also present a brief analysis of absorption feature statistics. Section 
4 summarizes our results and indicates some areas of future research.

\section{OBSERVATIONS AND THE ON-OFF METHOD}

\subsection{The CGPS 21-cm dataset}

The Canadian Galactic Plane Survey \citep{t03} is a multi-wavelength study at 1 arc-minute resolution 
from Galactic longitude 74\degr to 147\degr, and latitude -3.6\degr
to 5.6\degr. The observations for the 21 cm dataset were carried out with the Dominion Radio 
Astrophysical Observatory (DRAO) Synthesis Array, near Penticton, Canada 
\cite[for a description, see][]{ldbetal00}. The array consists of seven 
9-metre antennas. Four of the antennas are movable to yield a total of 12 configurations, or 144 baselines.
For the survey each field was observed in each position for 12 hours to obtain full uv-sampling.
Short spacing data were taken with the DRAO 26-m single dish telescope. At 21-cm the CGPS images have
a spatial resolution of 1\arcmin$\times$1\arcmin cosec($\delta$). The 
instrumental velocity resolution of 1.319 km/s is sampled every 0.824 km/s, covering LSR velocities from 
approximately 40 km/s to -160 km/s. The RMS noise of the emission in a single velocity channel at the field 
centres is about 3 K in low emission regions, and up to 6 K in high emission regions \citep{gthd00}. 
The continuum data are taken in
four 7.5 MHz bands (centred on 1406.9 MHz, 1413.8 MHz, 1427.4 MHz, and 1434.3 MHz).
A technical matter of importance for our work is the fact that a slightly different uv-taper is applied to 
the (continuum subtracted) \HI datacube than the continuum maps. We convolved the CGPS continuum maps to match 
the beam shapes exactly. For a detailed description of the data acquisition and reduction process see \cite{t03}.

\subsection{Compact Continuum Background Sources in the CGPS}

\begin{figure}
\plotone{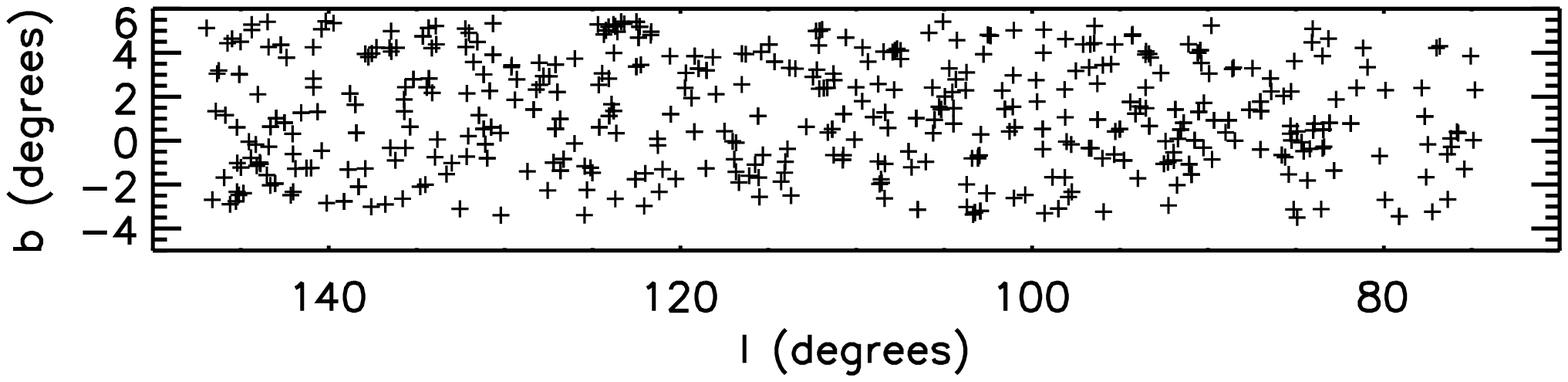}
\caption{Location of sources used for this work. \label{fig:sloc}}
\end{figure}

For this study we have identified 437 compact sources with a peak continuum brightness temperature higher than 
30 K in the CGPS, corresponding to integrated flux densities higher than 150 mJy.
The cut-off value of 30 K was chosen because the absorption signal to noise for sources 
below this temperature is almost always less than 1. This gives a source density of about 0.6 sources 
per square degree in the CGPS area. Figure~\ref{fig:sloc} shows the source locations. The sources 
provide a fairly uniform coverage of this section of the Galactic plane. Note that we 
do not exclude any sources that might be of Galactic origin. By inspecting the emission 
versus absorption velocity cut-offs we find about 15 potential Galactic sources.

\subsection{The On-Off Method}

Past emission-absorption studies typically used one on-source position and several off-source 
positions distributed around the source to approximate the "unabsorbed" spectrum 
(e.g. a hexagonal pattern, see \cite{cst88}).
For this work a different approach was taken, since we have a fully sampled, mosaiced 
dataset. First, the continuum sources were fitted with 2-D elliptical Gaussians. An average off-source emission 
brightness temperature, $T_{B}(v)$, was then extracted in a one arc-minute wide elliptical annulus, with 
its inner edge tracing the fitted ellipse in shape but scaled to be one FWHM from the centre. 
Most sources are unresolved, so the annuli are typically approximately one arc-minute (one beamwidth) 
away from the source centre. The RMS variations in $T_B(v)$ over the annulus give an estimate of the 
emission error. 
The off temperature is assumed to represent the emission on-source in the 
absence of continuum absorption. It is thus important to keep the separation between the on and off 
measurement directions as small as possible. At 1 kpc line of sight distance, one arc-minute corresponds 
to about 0.3 pc. 

An average optical depth ($\tau$) was computed from the on-source spectrum
using pixel values within one HWHM from the centre of the background source, and any pixels further 
away for which the emission temperature signal to noise was higher than 5. 
Under Local Thermal Equilibrium (LTE) the optical depth in any channel is then given by
\begin{equation} \label{eqn:tau}
\mathrm{exp}\left[-\tau(v)\right] = \left<1-\frac{T_{B} - T_{on}(x,y)}{T_{bg}(x,y)}\right>
\end{equation}
where $T_{bg}$ is the continuum brightness temperature of the background source and $T_{on}(x,y)$
is the continuum subtracted profile at the pixel location ($x,y$).
Typically about 10 pixels per channel are used in the average optical depth computation.
The spin temperature is derived by,
\begin{equation} \label{eqn:Ts}
T_s(v) = \frac{T_{B}(v)}{1-e^{-\tau(v)}}
\end{equation}
Note that for large optical depth, $T_{B}-T_{on}$ approaches $T_{bg}$. Due to noise in the data the 
optical depth can diverge. This situation was typically encountered for at least a few channels 
in any given spectrum. Since this corresponds to completely optically thick gas, 
we set $\mathrm{exp}(-\tau)$ and $T_s(v)$ equal to zero and $T_B(v)$ respectively in these cases.

To test our on-off method we inserted absorption from artificial background
sources into the data cubes.  The artificial continuum sources were Gaussian
shaped to match the synthesized beam and had peak brightness temperatures
ranging from 30 K to 300 K.  The line emission brightness temperature in each
channel was modified at the position of the artificial source in proportion
to the background source intensity and the \HI brightness in that channel at
the source position.   The resulting absorption spectra were extracted using
the on-off method and compared to the known input profiles. The agreement
between the two was consistently well within one sigma for both the spin
temperatures and optical depth spectra. We also compared spectra common to this study 
and \cite{dkgh83}. In general the agreement is good, but in noisy channels they sometimes differ
by several $\sigma$.

\section{RESULTS AND INTERPRETATION}

\begin{figure}
\epsscale{0.9}
\plotone{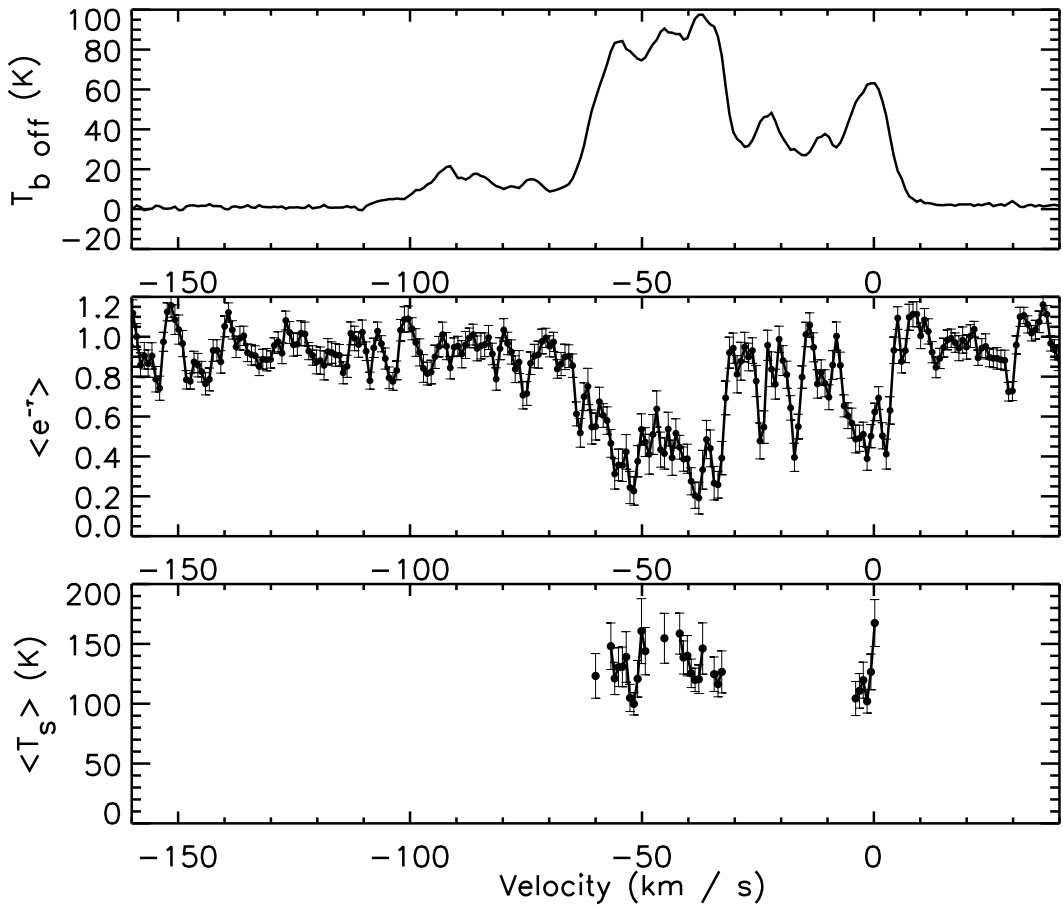}
\caption{Spectra for Maffei 2 ($T_{bg} = 32$ K). 1 $\sigma$ error bars are shown. \label{fig:spec01}}
\end{figure}

\begin{figure}
\epsscale{0.9}
\plotone{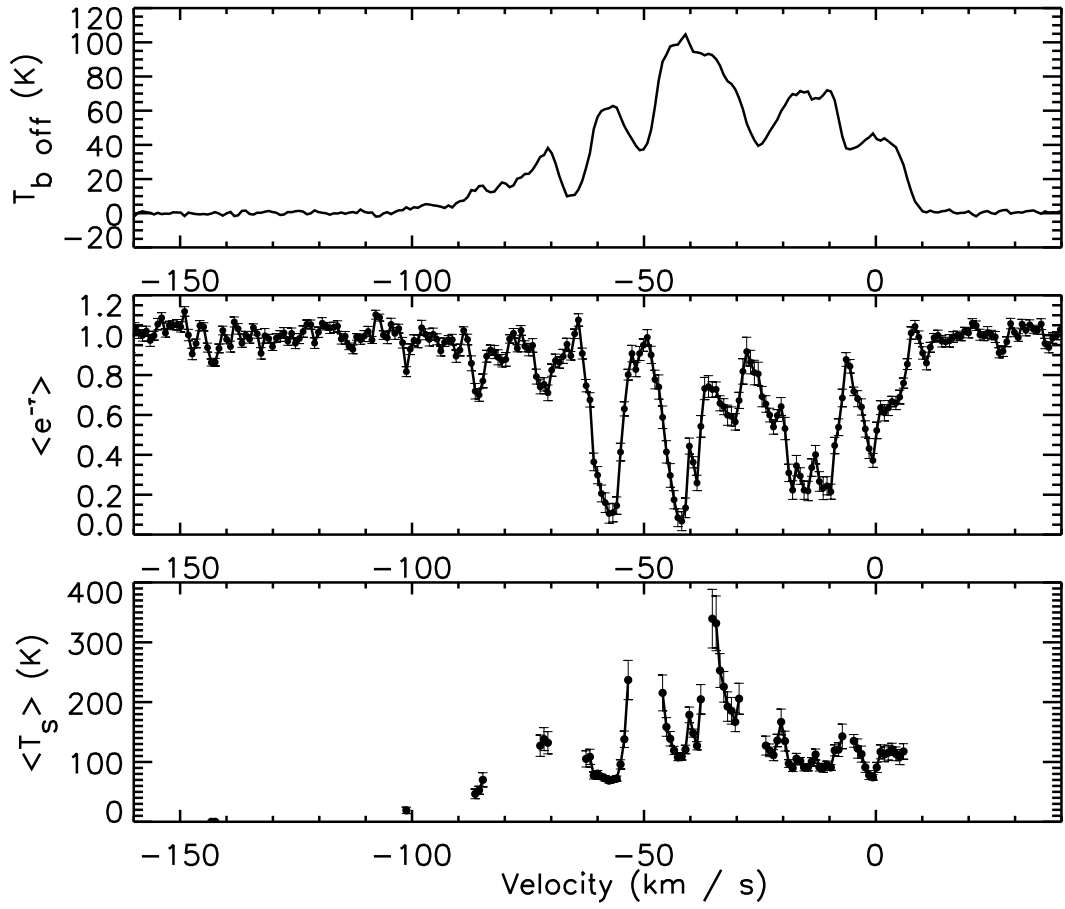}
\caption{Spectra for 87GB 032246.6+572847 ($T_{bg} = 77$ K). \label{fig:spec02}}
\end{figure}

Figures~\ref{fig:spec01} and \ref{fig:spec02} show two representative 
spectra, illustrating the quality of the data for two sources with very different continuum 
brightness temperatures. We do not show all the spectra in this paper due to their large number.
All the spectral data are available online ({\it{http://www.ras.ucalgary.ca/CGPS/products/}}) or
by request from S. Strasser. Some features are common to many spectra.
We can almost always identify the local (10 to -20 km/s) and Perseus arms (around -25 to -50 km/s) 
as broad peaks in emission, and in many cases also the outer arm (between -80 and -110 km/s). 
In general the absorption spectra show more structure and narrower line widths than the 
emission spectra. However this difference is less pronounced than for past, lower
angular resolution, studies. Some features follow the ``$T_s - \tau$ relation'' ($\log \tau$ 
decreases linearly with increasing $\log T_s$) quite nicely. This is attributed to 
the difference in absorption versus 
emission line widths, as the former is dominated by cool gas, and the latter by the warm component. 
In many cases however, the observed relation is more in the shape of 
a loop, where the spin temperature for a given optical depth on one side of the feature is systematically 
higher or lower than on the other side. Several recent studies confirm this relationship 
\citep[see e.g.][]{dmgg02}.

\subsection{The Spin Temperatures}

\begin{figure}
\epsscale{0.78}
\plotone{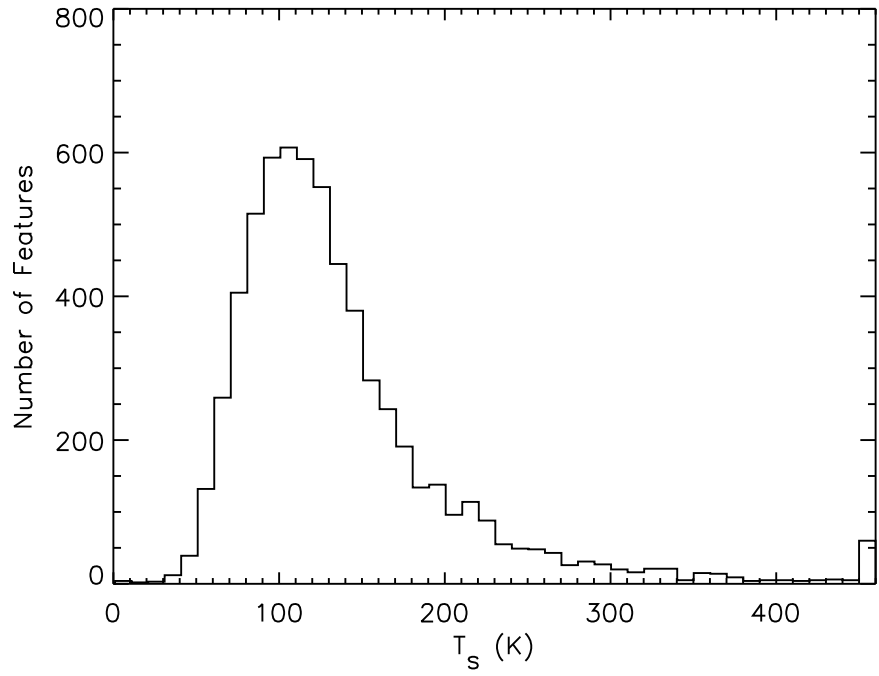}
\caption{The distribution of spin temperatures for all channels where $T_{on-off}$ S/N $>$ 5. The
last bin contains all channels with $T_s$ higher than 450 K.
\label{fig:Tshist}}
\end{figure}

The spin temperature ($T_s$) is the simplest temperature indicator we can compute. Under assumption of 
a single-component medium in equilibrium it is equal to the kinetic temperature of the gas 
\citep[e.g.][]{kh88}. The distribution of spin temperatures is given in figure~\ref{fig:Tshist} for
all channels with S/N$(T_{on-off}$) $>$ 5 (6318 channels in total). The median of the distribution is  
120 K, and 90\% of the values lie between 60 K and about 280 K. This is in rough agreement with 
previous studies \citep{dst78,mwkg82,pst82} and recent results for the southern Galactic
plane \citep{dmgg02}. 

It is generally believed though that the neutral hydrogen has
at least two components, the cold neutral medium (CNM) and the warm neutral medium (WNM).
The spin temperature measures a harmonic mean of the CNM and WNM, weighted by the column density of each 
component. It therefore only provides us with an upper limit to the true cold component temperature and a lower 
limit to the warm component temperature. This is confirmed by more recent analyses that 
fit the CNM and WNM separately and find lower cold component \HI temperatures 
than the associated spin temperatures. By fitting profiles with Gaussian functions \cite{ht02,ht02b} find cold 
component temperatures in the range from about 10 K to 75 K. \cite{dmgg02} use a novel technique of 
fitting the emission and absorption and find a range for the cold temperature between 20 K and 110 K, much lower
than their spin temperatures.

\subsection{A 2-Component Model} \label{sec:2c}

We have applied a two-component model to 
the data extracted from the CGPS. Our approach is similar to a model applied to the Galaxy and M31 by
\cite{bw92}. We use the transfer equation for a case where two components contribute to the emission 
brightness temperature: 
\begin{equation} \label{eqn:multi}
T_{B}(v) = T_w\left(1-e^{-\tau_w(v)q}\right) + T_c\left(1-e^{-\tau_c(v)}\right)e^{-\tau_w(v)q} + 
   T_w\left( 1-e^{-\tau_w(v)(1-q)}\right)e^{-\left[\tau_c(v) + \tau_w(v)q\right]}
\end{equation}
Here we assume that the gas has a cold component at temperature $T_c$ and optical depth $\tau_c(v)$ and a 
warm component at temperature $T_w$ with optical depth $\tau_w(v)$. A fraction $q$ of the warm
gas is in front of the cold gas at every velocity. The first term in equation~\ref{eqn:multi} 
represents WNM emission in front of the CNM. The second term accounts for the CNM emission, absorbed 
by the WNM in front. The third term represents the WNM behind the cold component, which is 
absorbed by both the CNM and WNM foregrounds. Since $\tau \propto N/T$, we further 
assume that the absorption due to the WNM is much smaller than that due to the CNM and that the 
WNM does not absorb the CNM. For 
$\tau_w(v) \ll \tau_c(v)$ and $\tau_w(v) \ll 1$ we then derive
\begin{equation} \label{eqn:2comp}
T_B(v) = T_0(v)e^{-\tau_c(v)} + T_{\infty}(v)\left(1 - e^{-\tau_c(v)}\right) 
\end{equation}
where $T_0 = T_w\tau_w(v)$ and $T_{\infty} = T_w\tau_w(v)q + T_c$ are the limits of $T_B$ as $\tau_c$
goes to zero and infinity respectively. Note that, even though these limiting temperatures 
are functions of velocity, we treat them as constants by assuming that $\tau_w(v)$ is approximately constant.
We make this assumption because we fit the data for all sources at once, 
hence averaging out the velocity dependence.

\begin{figure}
\plotone{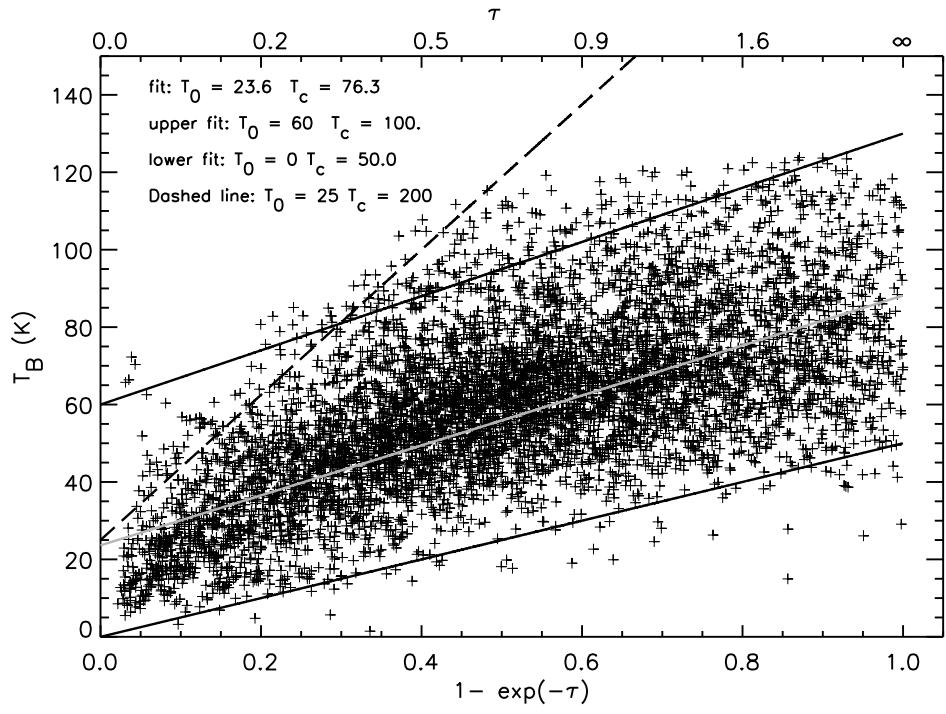}
\caption{$T_{B}$ versus 1-e$^{-\tau}$ with least squares fit and envelopes. The optical depth goes to zero
on the left, and approaches infinity on the right. 
\label{fig:2cfig01}}
\end{figure}

In figure~\ref{fig:2cfig01} a scatter plot of $T_{B}(v)$ versus $1-e^{-\tau_c(v)}$ for all sources is 
displayed. On these axes $T_0$ and $T_{\infty}$ are the values of the 
$T_B(v)$ at $1-e^{\tau_c(v)}$ equal to 0 and 1 respectively. Every data point represents one velocity channel. 
Only points with $T_{on-off}$ S/N higher than 5 are plotted. The centre line is a least 
squares fit of equation~\ref{eqn:2comp} to the data, yielding $T_0$ = 24 K and $T_{\infty}$ = 88 K.
We cannot explicitly solve for $T_w$, only the factor $T_{\infty} = T_w\tau_w$. 
Statistically, however, we expect an average value of $q$ = 0.5, and assuming this 
we can solve for the cold component temperature. This gives $T_c$ = 76 K. Varying $q$ gives an estimate of 
the uncertainty in the average cold component temperature. $T_c$ goes to 88 K and 64 K for $q$ equal to
0 and 1 respectively.

We interpret the scatter in the fit as being due to a range of $T_c$ and $T_0 = T_w\tau_w$. 
In figure~\ref{fig:2cfig01} we show visually selected upper and lower envelopes 
with the form of equation~\ref{eqn:2comp} (solid lines). $T_0$ ranges from 0 K to 60 K. 
Points on the lower envelope, which has $T_0=0$, thus represent regions where there is no warm component, 
and as we move toward the upper envelope, $T_c$ and $T_{\infty}$ increase. From the envelopes
we find $T_c$ ranging between 50 K and $100 \pm 30$ K (the upper limit is for $q$ = 0.5 and the uncertainty 
comes from varying $q$ between 0 and 1).

\begin{figure}
\plotone{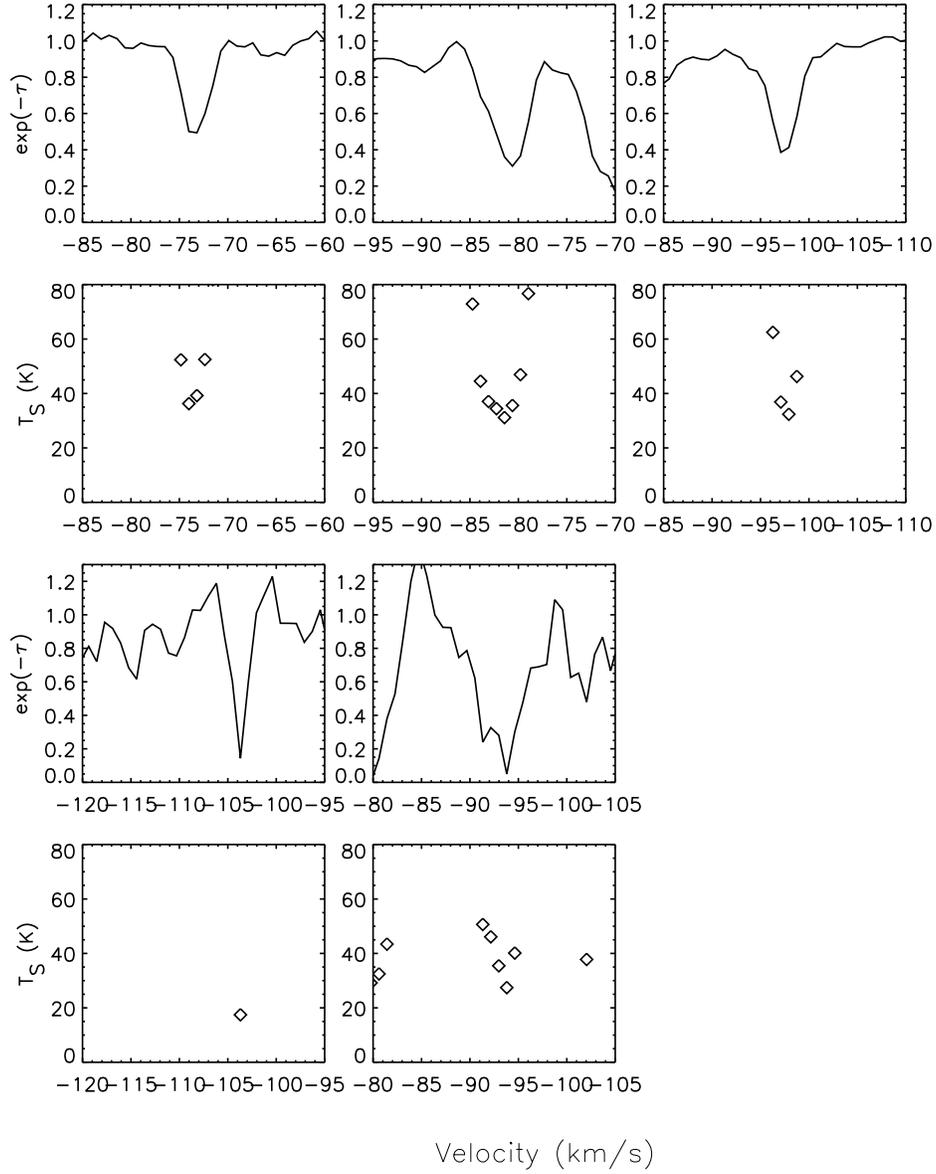}
\caption{The optical depth and spin temperature line profiles for the 5 identified cold features. The arrangement
from top left to bottom right is 4C +35.49, 3C 434.1, 87GB 020322.8+623159, MG4 J203647+4654, and 
87GB 231107.2+625244.
\label{fig:coldbox}}
\end{figure}

There are a few scattered points below the lower envelope. We analyzed these points in more detail to see 
if they could be due to a very cold neutral hydrogen component (below 40 K). We find that several well defined
(relatively unblended) features contribute most of the points. Table~\ref{tab:coldfea} summarizes these 
features. Selecting features with line centre $T_s$ 
below 40 K yields the same list. The optical depth and spin temperature line profiles for these 
features are shown in 
figure~\ref{fig:coldbox}. The last column in table~\ref{tab:coldfea} 
shows the cold temperatures obtained by fitting our two-component model to each 
feature region individually (for a $q$ = 0.5). Fitting individual features in this way is similar to the 
approach taken by \cite{dmss00}, except that we fit two components instead of one, and the fitting is done 
by least squares rather than by eye. The cold temperatures 
are confirmed by the two-component model. Except for the feature towards (85.089, +3.620), which has a 
computed $T_c$ of only 4 K (range of 0 K for $q=1$ to 13 K for $q=0$), all of the features have a 
$T_c$ of approximately 20 K, with a typical range between 10 K for $q=1$ and 30 K for $q=0$. 
Note that the velocities of these features place all of them quite far in the outer Galaxy  
(unfortunately making it unlikely that they show up as \HI self-absorption). 

While the model envelopes (figure~\ref{fig:2cfig01}) fit the data reasonably well, 
we can clearly see that $T_{off}$ increases more rapidly at small 
$\tau_c$ and flattens out at a maximum value of approximately 120 K. This peak brightness temperature is a 
well known observational fact for spiral galaxies. We are probably seeing two effects here: First, because 
galactic rotation spreads out the gas in velocity, only a certain amount of gas ever piles up at the same 
velocity limiting the maximum brightness temperature \citep{dmgg02}. Second, clearly if the optical depth 
in a line becomes much greater than 1, it will start to saturate, which creates a non-linear dependence 
of $T_B$ on $e^{-\tau}$. These two effects would explain why a simple radiation transfer, as 
modeled, does not fit the upper envelope better. The dashed line in figure~\ref{fig:2cfig01} 
shows a fit to the upper envelope below the ``break'' with $T_0$ = 25 K and 
$T_{\infty}$ = 212 K. For $q$ = 0.5 this gives a cold component temperature, $T_c$ = 200 K 
(with a range from 188 K to 212 K for $q$ equal to 1 and 0 respectively). If saturation is a dominant 
effect at high optical depth, then this line reflects what the state of the unsaturated gas is. The 
much higher $T_c$ value obtained by this fit, compared to fitting the whole range of $\tau$, is accounted 
for by the fact that the warmer the gas gets, the harder it is for it to reach high optical depth.

\cite{bw92}, applied a similar model to data 
for 87 sources covering a large range of Galactic latitude. They found best fit values of $T_c$ = 105 K and 
$T_0$ = 4 K, and envelope values from 54 K to 157 K for $T_c$ and from 2 K to 6 K for $T_0$. The inconsistency
between these and our results might arise primarily from the typically higher latitude of sources used 
in their study.

\subsection{Radial Variation of Physical Conditions}
\subsubsection{Temperature}

We examined the radial variation of the temperature by fitting the two-component model to bins in Galactocentric
radius. We used the rotation curve from \cite{bb93} to compute the Galactocentric radius corresponding to a given
velocity, assuming a rotation velocity of 220 km/s for the local standard of rest, and 8.5 kpc Galactic
centre to Sun distance. To find the radial variation of our model parameters, $T_c$ and $T_0$, we binned 
the data in 1 kpc intervals out to 16 kpc and beyond in 2 kpc intervals to 20 kpc. The bin sizes were chosen so
that every bin contains a large enough number of points to apply our model. We fitted the data for each 
bin in the same way as the whole dataset in section~\ref{sec:2c}, finding a best fit value and an upper 
and lower envelope.

\begin{figure}
\plotone{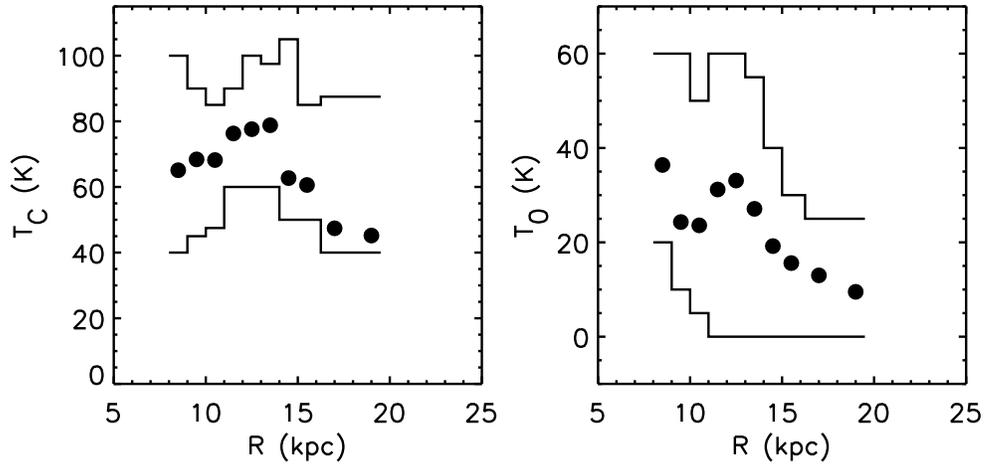}
\caption{Radial variation of the cold component temperature (left) and $T_w\tau_w$ (right). 
The lines show the values of $T_c$ and $T_w\tau_w$ derived from fitting the scatter envelopes. All values
plotted are for $q$ = 0.5.
\label{fig:2cradvar}}
\end{figure}

Figure~\ref{fig:2cradvar} shows the variation of $T_c$ and $T_0$ as a function of radius for $q$ = 0.5.
The dots show the fitted values, while the lines above and 
below indicate the temperature range given by the envelopes. The average cold component temperature rises 
from a local value of about 65 K to 80 K around 14 kpc and then drops off to 45 K in the outer Galaxy. 
The uncertainty introduced by $q$ is a maximum of about 20 K for the inner bins, and drops to around 10 K in
the outer bins (R $>$ 14 kpc). The lower 
envelope follows the same trend ranging from 40 K to 60 K, while the upper envelope depends only weakly 
on R ranging from 80 K to 100 K, with a minimum around 12 kpc and a peak at 14 kpc.

$T_0$ starts out at a maximum value of 36 K, and also has a local peak around 14 kpc before falling 
off to about 10 K at 20 kpc. These results indicate that warm gas is rare in the outer Galaxy. 
Of course the variation in $T_0$ could be due to a combination of 
changing warm temperature and opacity. Further, at 20 kpc a line of sight at latitude 5$^{\circ}$ 
is 1.8 kpc above the plane, and the Galactic warp could be as large as 4 kpc \citep{dl90}.
Especially the negative latitude lines of sight can thus look at quite a different height of the disk 
locally as compared to the far outer Galaxy. This could affect our results (our spin temperatures, however,
do not indicate any dependence on $z$).

\subsubsection{Volume Densities}

\begin{figure}
\plotone{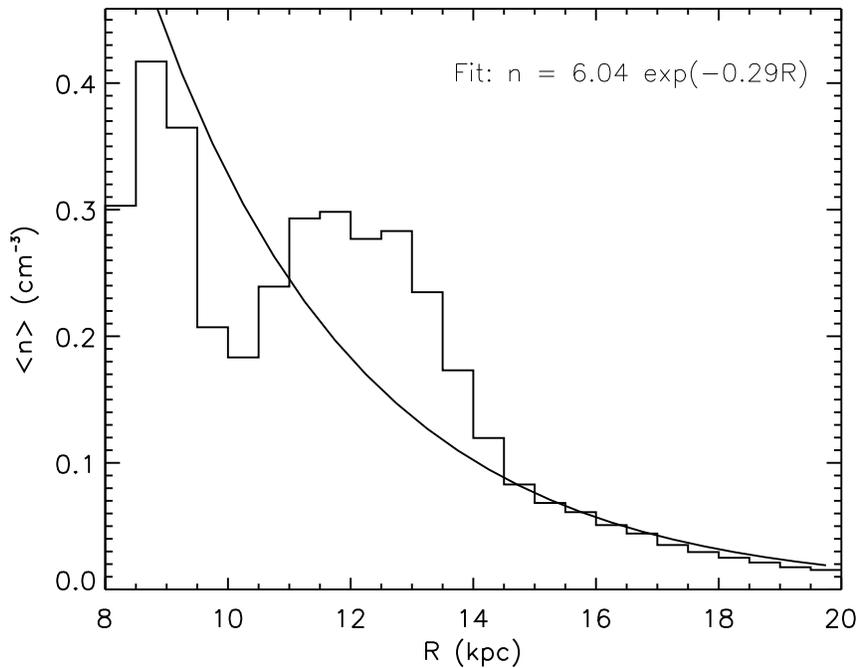}
\caption{Mean spatial density versus Galactocentric radius. The distribution is least squares fit with an exponential 
function.   
\label{fig:NvsR}}
\end{figure}

We have used the absorption and emission spectra to study the 
density of the \HI gas. For a gas at temperature $T_s= T_B\left(1-e^{-\tau}\right)$ the element 
of column density in a velocity channel of width $\Delta v$ is given by
\begin{equation} \label{eqn:N}
\Delta N(v) = \frac{C\tau(v)T_B(v)}{1-e^{-\tau}}\Delta v
\end{equation}
and represents the number of atoms per cm$^2$ in the velocity range $v$ to $v + \Delta v$. For $v$ in 
km/s, the constant $C$ is equal to $1.83\times10^{18}$ cm$^{-2}$K$^{-1}$(km/s)$^{-1}$. A spatial
density as a function of Galactocentric radius can be computed by dividing the line of sight into $R$ bins.
The density, $n(R)$ in any bin is given by,
\begin{equation} \label{eqn:nR}
n(R) = \frac{\sum_{\Delta V}N(v)}{\Delta L} \end{equation}
where $\Delta L$ is the line of sight distance through the bin and $\Delta V$ is the corresponding velocity 
range computed from \cite{bb93}. 
Figure~\ref{fig:NvsR} shows the mean density as a function of Galactocentric radius for all the sources.
A robust exponential least-squares fit of the form $y = Ae^{-Bx}$ to the distribution is shown, which 
characterizes the 
decrease quite well beyond 14 kpc, with values for $A = 6.04$ cm$^{-3}$ and $B = 0.29$ kpc$^{-1}$. 
We can see a local increase in the density around 12 kpc, due to the Perseus arm. 

\begin{figure}
\plotone{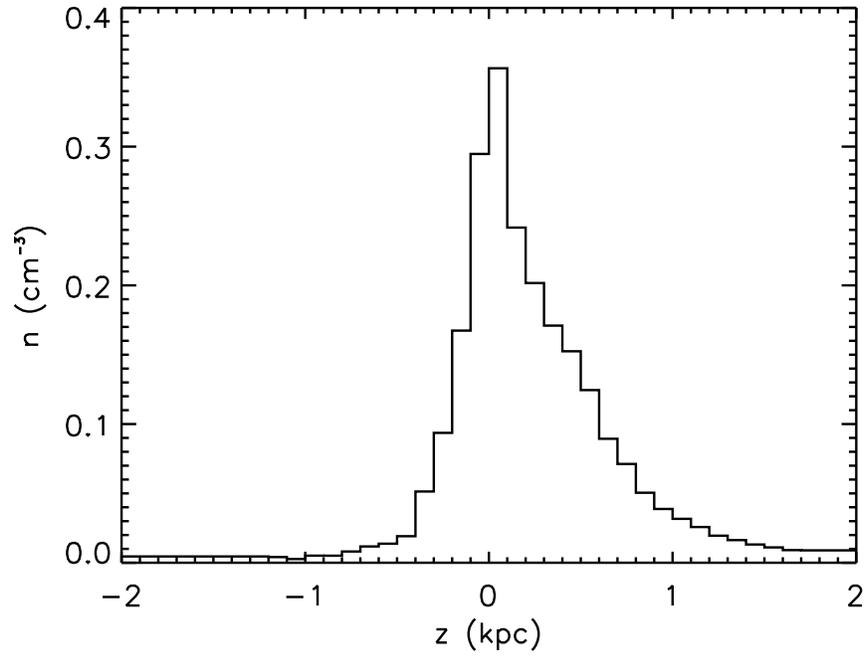}
\caption{Mean spatial density versus height above the Galactic plane. Note that the ``far'' solution was used 
for channels where a velocity ambiguity exists (inside the solar circle).
\label{fig:Nz}}
\end{figure}

We also examine the variation of density with height above the plane, $z$.
To derive a spatial density as a function of $z$, $n(z)$,  we sum the data in a given $z$ bin as in 
equation~\ref{eqn:nR}, but now $\Delta V$ refers to the velocity range contributing to the bin and $\Delta L$ 
is the line of sight range, as before.  
In figure~\ref{fig:Nz} we show the mean value of $n(v)$ as a function of $z$. The mean 
of the density drops off with scale heights of approximately 350 and 280 pc above and below the plane
respectively (found from exponential fits to the positive and negative sides of the distribution). The 
difference is due to the Galactic warp, which shows up as a density enhancement clearly visible at 
positive $z$ in figure~\ref{fig:Nz}. The scale heights correspond to a combination of the warm and 
cold components (averaged over the CGPS region).

\begin{figure}
\plotone{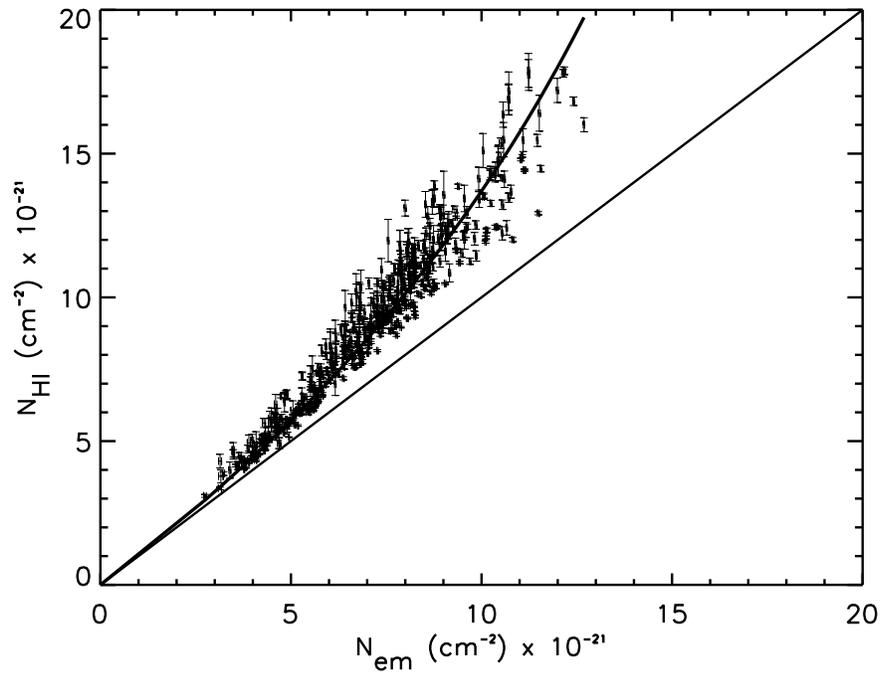}
\caption{The emission column density versus total column density (assuming a one phase medium) for all the sources. 
The lines show a 1:1 relationship and an empirical fit to the data (equation~\ref{eqn:Nfit}).
\label{fig:NvsN}}
\end{figure}

In the absence of information on absorption, \HI mass is often inferred from emission line data. The question of 
the ability to trace \HI gas by the emission alone is thus interesting to address. In deriving \HI mass from
emission the optically-thin case is assumed where equation~\ref{eqn:N} reduces to
\begin{equation} \label{eqn:Nem}
N_{em}(v) = C T_B(v)\Delta v, \end{equation}
where we have designated $N_{em}$ as the column density derived from emission alone.
Figure~\ref{fig:NvsN} shows the emission column density, $N_{em}$, versus the column density $N_{HI}$
given by absorption using equation~\ref{eqn:N}. By combining equation \ref{eqn:N} with 
equation~\ref{eqn:Nem} we can find a functional relation between $N_{HI}$ and $N_{em}$, 
\begin{equation} \label{eqn:Nfit}
N_{HI} = -A \ln \left(1 - \frac{N_{em}}{A}  \right) \end{equation}
where $A = C T_s \Delta v$. In reality many different spin temperatures contribute to each measured 
column density. However equation~\ref{eqn:Nfit} provides a convenient way to characterize the 
relationship between $N_{HI}$ and $N_{em}$ . By least squares fitting the parameter A, 
we find a value of $2.1 \times 10^{22}$ cm$^{-2}$; The corresponding fit is shown in 
figure~\ref{fig:NvsN} and well represents the data.

The average ratio of $N_{HI}/N_{em}$ is equal to 1.32 (or $N_{em}$ = 75\% of $N_{HI}$). At the highest 
column densities $N_{em}$ measures only about 65\% of the true column. \cite{dmgg02} find typical values 
for $N_{HI}$/$N_{em}$ of 1.4 to 1.6 in the inner Galaxy.\cite{dkgh83} derived a somewhat lower 
value of 1.2 for $N_{HI}/N_{em}$. The difference is probably due to our concentration of lines of sight very 
close to the Galactic plane, where the opacity is highest, so the difference between $N_{HI}$ and $N_{em}$ 
is the largest on average. \cite{ht02} clearly show this dependence of $N_{HI}/N_{em}$ on Galactic 
latitude in their figure 5, and their values close to the Galactic plane qualitatively agree with the 
results presented here.

\subsection{Correlation with CO}

Another of the CGPS data products are the reprocessed images from the 
FCRAO Outer Galaxy Survey of the $^{12}$CO J = 1 - 0 rotational transition 
at 2.6 mm \citep{hbs98}. The survey has an angular resolution of 1.67 arc-minutes, and covers 
the same velocity range as the 21-cm dataset at the same spectral resolution. 
CO is often used as a tracer for molecular hydrogen, as it exists in a similar physical environment. The 
accuracy with which it traces $H_2$, however, is still a subject of frequent debate. 
While the presence of CO correlates with $H_2$ or very cold atomic hydrogen, the reverse often does not 
seem to be the case \cite[e.g][]{rkh94,mh96}.

Recent studies \citep{gthd00,ht02b,kb01,kd03} are finding clear evidence for \HI below 30 K, where we expect most 
the hydrogen to be in molecular form . Our own 2-component analysis reveals several 
absorption features below 40 K. To check for the co-existence of \HI absorbing gas and molecular 
material we compared the \HI absorption spectra to the CO emission spectra towards the sources. 

\begin{figure}
\plotone{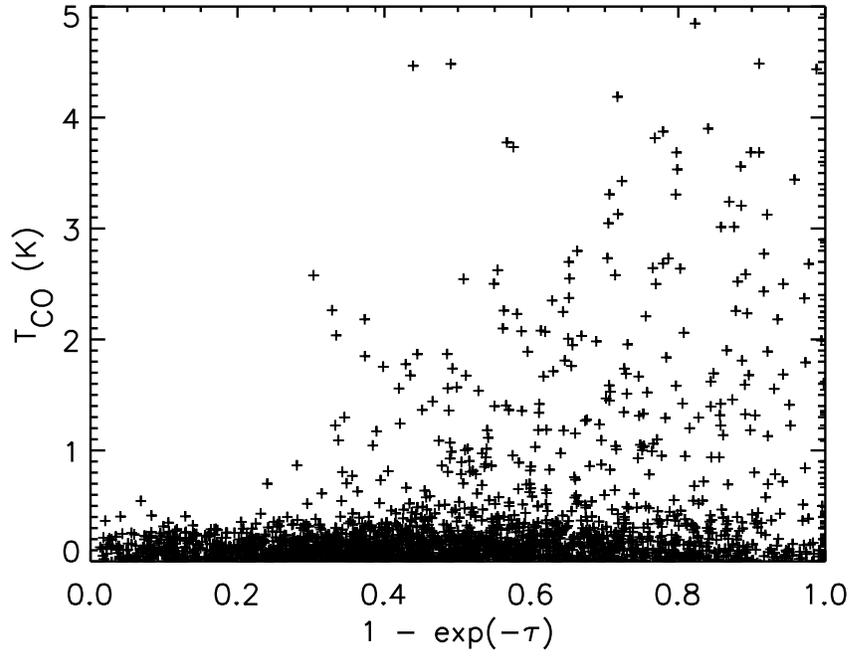}
\caption{CO emission brightness temperature versus $1 - e^{-\tau}$ for the 134 sources that are in the 
FCRAO survey field.
\label{fig:COcorr}}
\end{figure}

\begin{figure}
\plotone{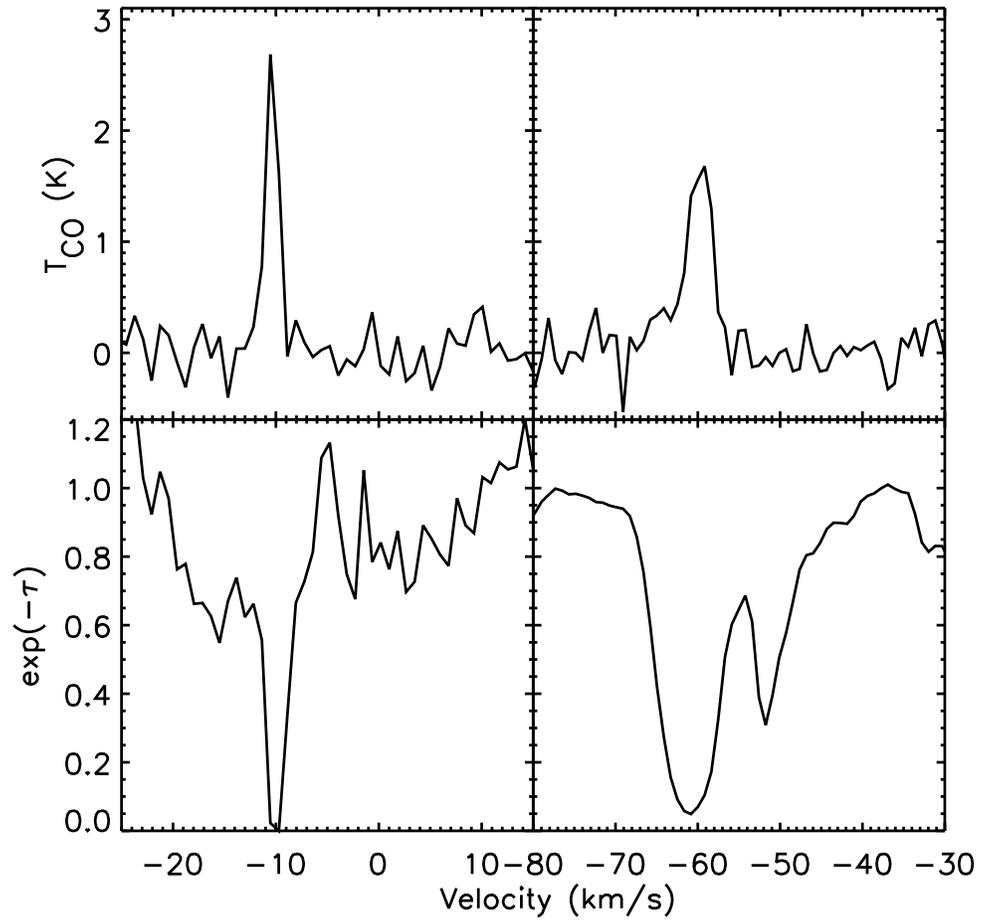}
\caption{Two example features that appear both in CO and optical depth. On the left 87GB 013134.9+650922,
on the right 3C 011.1.
\label{fig:COsamplespecs}}
\end{figure}

Figure~\ref{fig:COcorr} shows the CO emission brightness temperature 
versus the \HI optical depth for the 134 sources which are in the FCRAO survey area (from Galactic 
longitude 102$^\circ$ to 143$^\circ$ and all CGPS latitudes). It is evident that overall there is no correlation 
between the CO emission and \HI absorption. The CO noise level is $\sim 0.15$ K. There is 
a very clear lack of data points with CO emission brightness temperature above the noise level at
low \HI optical depth. This, and visual inspection of the spectra, show that in most 
cases where we do see CO emission, \HI absorption is also present. 
We find about 60 clear matches between \HI and CO features, and for approximately 20 of 
these, $e^{-\tau}$ goes to zero. Many such features are only a few channels wide both 
in CO and \HI optical depth. Figure~\ref{fig:COsamplespecs} shows two sample features.
The narrow feature in the left panel, illustrates a case in which $e^{-\tau}$ goes to zero. 
The panel on the right displays two broad \HI absorption lines, one of which shows a clear correlation with CO 
emission. 

Of the two cold features listed in table~\ref{tab:coldfea} for which we have CO data, one 
(112.107,+2.362) has associated CO emission. In general, however,
there is no preference for low spin temperatures in \HI channels with associated CO. As noted earlier 
though, if a cloud has multiple temperature components, the spin temperature is a harmonic mean, and 
therefore, even if the cold-component temperature is very low, the apparent spin temperature does not 
necessarily have to be low. So, while there is no overall correspondence between the CO and \HI features, 
the association of over 60 features 
is interesting, and a good indication that at least some of the atomic hydrogen is at very low 
temperatures. We plan to pursue this by modeling the individual features and by obtaining 
more observations at higher spectral resolution and signal to noise to be able to better quantify our 
current results.

\subsection{Feature Distribution and Properties}

The spatial density of absorption peaks is a simple tracer of cold ISM structures. An absorption feature 
in general corresponds to one cloud, or more if several lines are blended. The feature density 
(number of features per kpc) is thus a lower limit to the cloud density. We have developed an automatic 
absorption feature identification routine that finds peaks in the $T_{on-off}$ spectra. Such an 
identification is never completely unambiguous, as noise spikes have to be separated from real features. 
The criteria applied were chosen to agree with a 
visual inspection of the spectra. Features have to be stronger than 3 sigma in $T_{on-off}$ to be
identified. To determine the line of sight distance corresponding to a given radius interval we used the 
rotation curve from \cite{bb93}. Mathematically for any radial bin the feature density is given by, 
\begin{equation}
\mathrm{\#\,features\, /\, kpc} = \frac{\sum_{sources} \mathrm{\#\,features\,/\,kpc}}{N_{bin}}
\end{equation}
where $N_{bin}$ is the total number of sources contributing to any given bin.

\begin{figure}
\plotone{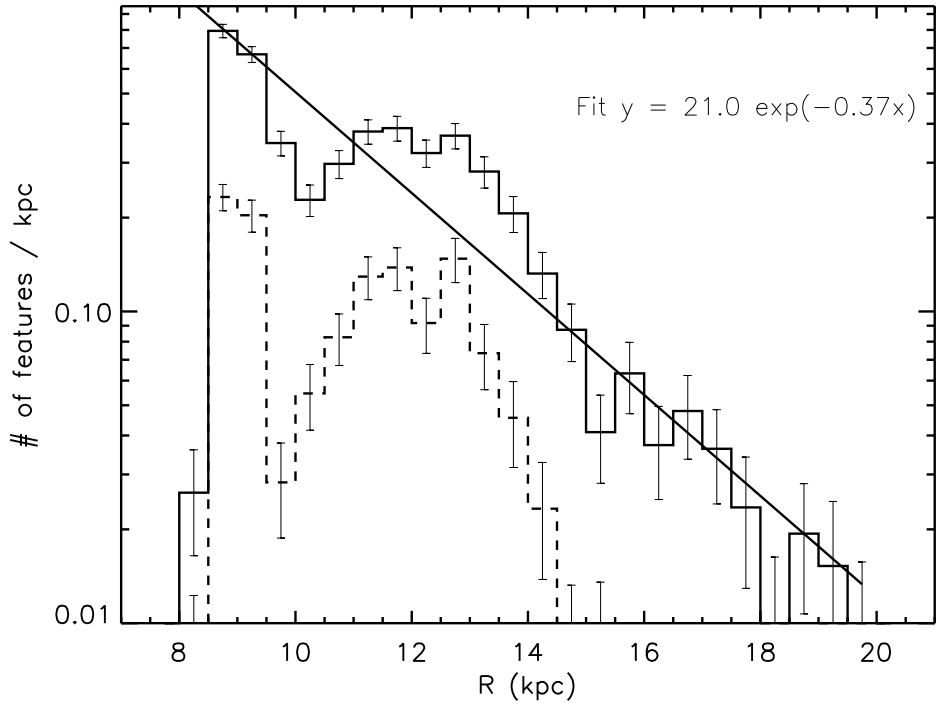}
\caption{Feature density versus Galactocentric radius for features with peak $\tau > 0.5$ (solid) and 
$\tau > 1.5$ (dashed).
\label{fig:feaR}}
\end{figure}

Figure~\ref{fig:feaR} shows the density of features as a function of Galactocentric
radius for two optical depth cut-off levels. The solid histogram is for all features with $\tau > 0.5$, while
the dashed histogram is for $\tau > 1.5$. The error bars indicate the standard deviation in the bin 
divided by the square-root of the number of sources contributing to each bin. 
We show an exponential least-squares fit to the distribution for $\tau > 0.5$ which is given by 
19.7e$^{-0.36R}$ features/kpc, and which describes the distribution quite well. 
As the histogram for $\tau > 1.5$ illustrates, higher optical depth features are clearly 
concentrated in the local and Perseus arms and disappear entirely beyond 16 kpc. 
\cite{gd89} found the feature density for all peak values 
of $\tau$ to have an average value of about 0.42 $\pm$ 0.08 kpc$^{-1}$ from R = 0 to 4.25 kpc and 
0.8 $\pm$ 0.1 from R = 4.25 kpc to 8.5 kpc which agrees with our results. The feature density thus seems to
reach a peak around the solar neighbourhood and decreases both in the inner and outer Galaxy. 

We have to keep in mind though that as the line of sight distance increases, a given change in velocity becomes
a larger and larger change in line of sight distance due to the rotation curve. While 1.32 km/s (our spectral 
resolution) locally corresponds to about 200 to 300 pc (using the rotation curve to convert from 
velocity to distance), at -100 km/s the same velocity corresponds to scales from 200 pc to a few kpc 
(depending on longitude). Thus, if the dominant scale of structure in the \HI is on the order of
a 100 pc rather than several hundred parsecs to kiloparsecs, then this observed radial dependence could at 
least partially be an observational effect. As we look further and further away along the line of sight, 
we lose clouds in our identification due to the decreasing spatial resolution. 
In the outer Galaxy, an increase in the line of sight distance always corresponds to an increase in 
Galactocentric radius. Thus, since our data is largely in the outer Galaxy, the observed effect will give 
a drop in the number of features.

\section{CONCLUSIONS}

We have presented a new emission-absorption survey of the Galactic Plane at 21-cm with an unprecedented source
density of about 0.6 sources per square degree. In agreement with previous studies we find spin temperatures
between 40 K and 300 K. Due to the difficulty in interpreting the spin temperature we apply a simple
2-component model to the data. We find $T_c$ and $T_w\tau_w$ ranging from approximately 50 K to 
80 K, and 0 K to 60 K respectively. Both $T_c$ and $T_w\tau_w$ show a peak at about 14 kpc and
a decrease in the outer Galaxy. 

The derived density of neutral atomic hydrogen varies roughly as e$^{-0.3R}$, with Galactocentric 
radius (in kpc). However directly interpreting
this as a drop in gas density with radius is complicated by the presence of the Galactic warp and the changing
scale height visible in the $z$ profile of N. Comparing the column density traced by emission to 
the total column we find that on average the emission traces about 75\% of the total \HI mass, and at the 
highest optical depths only about 65\%. 

We looked for evidence of very cold hydrogen in the Galaxy in our two-component model results and the spin
temperature. This yielded a list of 5 features, with $T_s < 40$ K and lying below the two-component 
model lower envelope.
Further a comparison to CO shows that even though there is no overall correlation between CO emission and 
\HI absorption, about half the spectra show at least one \HI absorption feature with clearly associated CO 
emission. 

This study is a first look at the \HI absorption in the CGPS. We hope to continue this analysis by 
incorporating the Southern and the VLA Galactic Plane Surveys \citep{mgdghw01,t03}, giving us 
more information on the inner Galaxy, and the difference between the southern and northern parts of the 
Galactic plane. While the 2-component model applied here gives us some interesting results, 
it is clear that a more detailed feature by feature analysis could reveal much new information. 
This will also be pursued further.

\acknowledgements

We wish to thank J.M. Dickey, S.J. Gibson, and C. Heiles for helpful comments and discussions. 
Thanks also to J.C. Brown for her help with IDL and to L.A. Higgs for modifying the DRAO software 
for our purposes.

\clearpage


\clearpage

\begin{deluxetable}{lcccccccc}
\tablecaption{Cold Spin Temperature Features \label{tab:coldfea}} 
\tablewidth{14.5cm}
\tabletypesize{\footnotesize}
\tablehead{
\colhead{Source Name} & \colhead{$l$} & \colhead{$b$} & \colhead{v} & 
\colhead{$T_s$} & \colhead{$\Delta T_s$} & \colhead{$\tau$} & \colhead{$T_{B}$} & \colhead{$T_c$} \\
  & \colhead{(deg)} & \colhead{(deg)} & \colhead{(km/s)} & 
\colhead{$(K)$} & \colhead{$(K)$} & \colhead{} & \colhead{$(K)$} & \colhead{$(K)$}
}
\startdata
4C+35.49 & 76.366  &  -2.674 &  -74.02  &  36  &  3   &  0.71  &  18 & 20 \\    
MG4 J203647+4654 & 85.089  &  3.620  &  -103.70 &  18  &  2  &  1.94  &  15 & 4 \\       
3C+434.1 & 94.112	&  1.223  &  -81.47  &  31  &  2  &  1.01  &  20 & 20 \\      
87GB 231107.2+625244 & 112.107 &  2.362  &  -93.80  &  27  &  2  &  3.00  &  26 & 19 \\      
87GB 020322.8+623159 & 131.459 &  1.157  &  -97.93  &  32  &  2  &  0.89  &  19 & 22 \\    
\enddata
\tablecomments{The values quoted are for the lowest $T_s$ point in each feature. $\Delta T_s$ is 
the uncertainty in $T_s$.}
\end{deluxetable} 

\clearpage

\end{document}